\documentclass[]{iac}
\usepackage{titlesec}
\usepackage{authblk}
\usepackage[labelsep=period]{caption}
\usepackage{multirow}
\usepackage{makecell}
\usepackage{tabu, booktabs}
\usepackage[table,x11names,dvipsnames,table]{xcolor}
\usepackage{float}
\usepackage{subcaption}
\usepackage{graphicx}
\usepackage{mathtools}

\usepackage{flushend}
\usepackage{stfloats}
\usepackage{url}
\usepackage{placeins}
\usepackage{multicol}
\titlespacing{\section}{0pt}{*1.2}{*.1}
\titlespacing{\subsection}{0pt}{*1.2}{*0.1}

\newtagform{brackets}{(}{)}
\usetagform{brackets}

\newcommand*{\vectornorm}[1]{\left\|#1\right\|}

\newcommand{\listheading}[1]{\noindent\textbf{#1}\par\nobreak\vspace{0.1\baselineskip}}

\apptocmd{\thebibliography}{\small\setlength{\itemsep}{0pt}\setlength{\parsep}{0pt}}{}{}

\newcommand{\ResultMissionDays}{381.3}
\newcommand{\ResultMissionDaysPrecise}{381.3148}
\newcommand{\ResultOrbitDaysPrecise}{381.3035}
\newcommand{\ResultCaptureDays}{184.1}
\newcommand{\ResultCaptureBracketDays}{183.7--184.1}
\newcommand{\ResultSMAKm}{3{,}862}
\newcommand{\ResultEcc}{0.0867}
\newcommand{\ResultInclinationDeg}{72.61}
\newcommand{\ResultCThree}{-1.269}
\newcommand{\ResultPeriluneKm}{1{,}790}
\newcommand{\ResultApoluneKm}{2{,}459}
\newcommand{\ResultDeltaV}{839.8}
\newcommand{\ResultThrustDays}{108.00}
\newcommand{\ResultFuelKg}{0.361}
\newcommand{\ResultMassKg}{12}
\newcommand{\ResultDryMassKg}{9}
\newcommand{\ResultPropellantKg}{3}
\newcommand{\ResultIsp}{2800}
\newcommand{\ResultExhaustKms}{27.47}
\newcommand{\ResultNominalThrustmN}{1.2}
\newcommand{\ResultEffectiveThrustmN}{1.08}
\newcommand{\ResultCircularizationArcs}{244}
\newcommand{\ResultCircularThrustDays}{74.09}
\newcommand{\ResultCircularDeltaV}{576.1}
\newcommand{\ResultCaptureBlockSMA}{27{,}721}
\newcommand{\ResultCaptureBlockEcc}{0.683}
\newcommand{\ResultPreFlybyStart}{0.08}
\newcommand{\ResultPreFlybyEnd}{2.06}
\newcommand{\ResultPreFlybyDuration}{1.980}
\newcommand{\ResultPreFlybyDeltaV}{15.4}
\newcommand{\ResultTransitionStart}{129.99}
\newcommand{\ResultTransitionEnd}{136.63}
\newcommand{\ResultTransitionDuration}{6.636}
\newcommand{\ResultTransitionDeltaV}{51.6}
\newcommand{\ResultCaptureStart}{169.65}
\newcommand{\ResultCaptureEnd}{194.95}
\newcommand{\ResultCaptureDuration}{25.300}
\newcommand{\ResultCaptureDeltaV}{196.7}
\newcommand{\ResultCircularStart}{199.88}
\newcommand{\ResultPreTransitionCoast}{127.93}
\newcommand{\ResultTransitionCoast}{33.018}

\newcommand{\IACfrontmatter}{%
\begingroup
\normalsize
\setlength{\baselineskip}{11.5pt}
\setlength{\parindent}{0pt}
\setlength{\multicolsep}{0pt}
\begin{multicols}{2}
\raggedright
\listheading{Nomenclature}

$a$: Semi-major axis, km\\
$C_3$: Characteristic energy, km$^2$/s$^2$\\
$e$: Orbital eccentricity\\
$F$: Thrust magnitude, N\\
$g_0$: Reference gravitational acceleration, m/s$^2$\\
$I_{\mathrm{sp}}$: Specific impulse, s\\
$m$, $m_0$: Modeled and initial mass, kg\\
$m_{p,\mathrm{eq}}$: Ideal propellant equivalent, kg\\
$s(t)$: Binary thrust state\\
$\mathbf{r}$, $r$: Position vector and its norm, km\\
$\rho$: Lunar-relative distance, km\\
$R_{\mathrm{M}}$: Lunar reference radius, km\\
$t$, $t_f$, $t_{\mathrm{end}}$: Time and endpoint epochs\\
$\Delta t$: Burn or coast duration\\
$\Delta V$: Integrated velocity increment, m/s\\
$\mu$: Gravitational parameter, km$^3$/s$^2$\\
$\nu$: True anomaly\\
$\mathbf{v}$: Velocity vector, km/s\\
$v_e$: Exhaust velocity, km/s\\
$\hat{\mathbf{d}}$: Unit thrust direction\\
$\mathbf{a}$: Acceleration vector, km/s$^2$\\
$\boldsymbol{\theta}$: Targeting parameter vector
\par\vspace{4pt}
\listheading{Acronyms/Abbreviations}

AoP: argument of periapsis\\
CAD: computer-aided design\\
CH-1: Chandrayaan-1\\
CubeSat: cube satellite\\
CZL: coronal and zodiacal light\\
ECI: Earth-centered inertial\\
Ecc: eccentricity\\
EP: electric propulsion\\
GMAT: General Mission Analysis Tool\\
GRAIL: Gravity Recovery and Interior Laboratory\\
GTO: geostationary transfer orbit\\
IAC: International Astronautical Congress\\
IAF: International Astronautical Federation\\
Inc: inclination\\
JPL: Jet Propulsion Laboratory\\
LEO: low Earth orbit\\
LHG: lunar horizon glow\\
LRO: Lunar Reconnaissance Orbiter\\
MCI: Moon-centered inertial\\
NASA: National Aeronautics and Space Administration\\
RAAN: right ascension of the ascending node\\
RMAG: position-vector magnitude\\
SLS: Space Launch System\\
SMA: semi-major axis\\
SMART-1: Small Missions for Advanced Research in Technology\\
SQP: sequential quadratic programming\\
TA: true anomaly\\
VNB: velocity-normal-binormal
\par
\end{multicols}
\endgroup
}

\makeatletter
\apptocmd{\@maketitle}{\IACfrontmatter}{}{}
\makeatother

\begin{document}

\IACpaperyear{20}
\IACpapernumber{C1,4,8,x55549}
\IACconference{71}
\IAClocation{The CyberSpace Edition, 12-14 October 2020}
\IACcopyrightA{2020}{International Astronautical Federation (IAF)}

\title{\textbf{LOW-THRUST TRAJECTORY OPTIMIZATION FOR CUBESAT LUNAR MISSION: HORYU-VI}}

\author[a,*]{Omer Burak Iskender}
\author[a]{Keck Voon Ling}
\author[b,a]{Mengu Cho}
\author[b]{Sangkyun Kim}
\author[b]{Necmi Cihan Orger}

\affil[a]{Nanyang Technological University, Singapore, iske0001@e.ntu.edu.sg}
\affil[b]{Kyushu Institute of Technology, Kitakyushu, Japan}
\affil[*]{Corresponding author}
\date{}
\setcounter{Maxaffil}{0}
\renewcommand\Affilfont{\itshape\small}
\abstract{\centerline{\textbf{Abstract}}\\ A three-stage low-thrust trajectory is investigated for the HORYU-VI CubeSat lunar mission: pre-flyby deceleration, lunar capture, and orbit circularization. Impulsive reference trajectories initialize finite-burn targeting in the General Mission Analysis Tool (GMAT). The archived case uses constant \ResultMassKg~kg spacecraft mass and \ResultEffectiveThrustmN~mN effective thrust. The first reported bound lunar state occurs at day~\ResultCaptureDays. At the end of the \ResultMissionDays-day record, the osculating lunar orbit has semi-major axis \ResultSMAKm~km and eccentricity \ResultEcc. Integrating the reported thrust state yields an estimated velocity increment of \ResultDeltaV~m/s over \ResultThrustDays~days of thrust. At the configured \ResultIsp~s specific impulse, the ideal rocket-equation propellant equivalent is \ResultFuelKg~kg; mass depletion was disabled in the propagation. The record demonstrates lunar capture and partial circularization. Completion of the intended 100~km science orbit, coupled mass-depleting propagation, and uncertainty analysis remain necessary to assess operational feasibility.\\}
\IACkeywords{Electric Propulsion}{Trajectory Optimization}{Low Thrust}{CubeSat}{Lunar Mission}{HORYU-VI}
\maketitle

\clearpage
\section{Introduction}

CubeSat platforms have rapidly expanded their reach from Low Earth Orbit (LEO) to cislunar and interplanetary space, offering cost-effective spacecraft for scientific and technological objectives~\cite{poghosyan2017cubesat,folta2016lunar,klesh2018marco,hardgrove2015lunar,cohen2015lunar}. Lunar science has emerged as a primary focus for next-generation CubeSat missions.

The HORYU-VI mission is an international CubeSat initiative to investigate the lunar dust environment through multi-epoch imaging of the Lunar Horizon Glow (LHG)~\cite{orger2020horyu}. The project is a collaboration among Kyushu Institute of Technology (Japan), Nanyang Technological University (Singapore), Sapienza University of Rome (Italy), California Polytechnic State University (USA), University of Colorado Boulder (USA), and Aliena Pte.\ Ltd.\ (Singapore). The 12U CubeSat platform carries dedicated imaging instruments for LHG observations from a low-altitude lunar orbit.

The LHG (an anomalous brightness above the lunar terminator exceeding the coronal and zodiacal light baseline) was first observed during the Apollo era~\cite{mccoy1974evidence,mccoy1976photometric,rennilson1974surveyor,criswell1973horizon}. Lunar dust poses hazards to both crewed missions (respiratory risks, contamination, long-term health effects)~\cite{harris1972apollo,gaier2005effects,james2009pulmonary,linnarsson2012toxicity} and robotic systems (optical degradation, thermal property alterations, electrical discharges)~\cite{gaier2007effects,o2011review,o2018paradigm}. Characterizing the near-surface dust population from orbit remains an open challenge in lunar science~\cite{severny1975measurements,glenar2011reanalysis,colwell2007lunar,feldman2014upper,barker2019searching}.

For this spacecraft, a nominal combined thrust of \ResultNominalThrustmN~mN produces an effective acceleration of approximately $9\times10^{-5}$~m/s$^2$ at the modeled wet mass. Consequently, substantial velocity changes accumulate over extended thrust arcs. Electric propulsion (EP) provides high exhaust velocity and enables long-duration low-thrust maneuvers within a limited propellant allocation. The trajectory design must account for gravitational dynamics during both thrust and coast intervals.

This paper addresses the trajectory design challenge for HORYU-VI. The launch opportunity was envisioned as a secondary payload on NASA's Space Launch System (SLS) under the Artemis program. Upon separation, the CubeSat enters a high-eccentricity orbit ($e = 0.9667$) with sufficient energy for lunar flyby and subsequent heliocentric escape if left uncontrolled. Converting this trajectory into a stable lunar orbit requires precise multi-phase low-thrust maneuvers within the severe propulsion constraints of a CubeSat platform.

\section{Trajectory Options}

Table~\ref{tab:traj_options} summarizes established lunar transfer methods~\cite{jpl2013}. The SMART-1 mission pioneered low-thrust cislunar transfers using a Hall-effect thruster, requiring $\sim$14~months to reach lunar orbit from GTO~\cite{racca2009smart}. Its orbit-raising, lunar-capture, and circularization sequence provides operational heritage for HORYU-VI. The present work adopts a similar phased decomposition but differs in that the initial orbit is a high-energy flyby trajectory rather than GTO.

\begin{table*}[!t]
\centering
\caption{Lunar transfer methods with typical duration and representative missions~\cite{jpl2013}.}
\label{tab:traj_options}
\rowcolors{1}{}{gray!10}
\resizebox{\textwidth}{!}{%
\begin{tabular}{@{}lcll@{}}
\toprule
Transfer Type        & Typical Duration & Benefits                   & Example               \\ \midrule
Direct, conventional & 3--6 days        & Well known, quick          & Apollo, LRO           \\
Direct, staging      & 2--10 weeks      & Quick, many launch days    & Clementine, CH-1      \\
Direct to lunar L1   & 1--5 weeks       & Staging at L1              & Few to date           \\
Low-thrust           & Many months      & Low fuel, many launch days & SMART-1               \\
Low-energy           & 2.5--4 months    & Low fuel, many launch days & Hiten, GRAIL \\ \bottomrule
\end{tabular}%
}
\footnotesize{LRO: Lunar Reconnaissance Orbiter; CH-1: Chandrayaan-1; SMART-1: Small Missions for Advanced Research in Technology.}\\
\end{table*}

\section{Mission Architecture}

\begin{figure}[!htbp]
    \centering
    \includegraphics[width=0.49\textwidth]{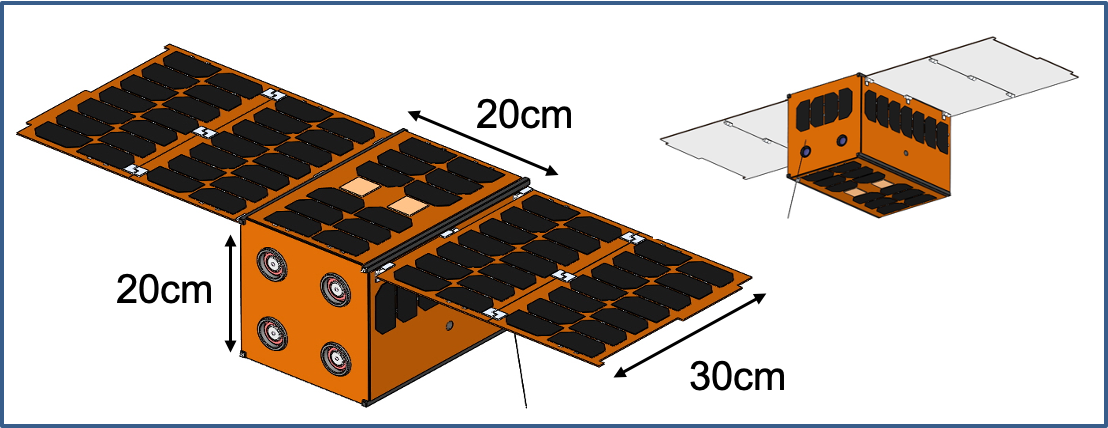}
    \caption{CAD design of HORYU-VI CubeSat.}
    \label{fig:design}
\end{figure}

The HORYU-VI spacecraft design uses a 12U CubeSat with four Hall-effect thrusters. The reference configuration for the numerical case has dry mass \ResultDryMassKg~kg and a \ResultPropellantKg~kg propellant allocation, giving a constant modeled mass of \ResultMassKg~kg. The nominal combined thrust is \ResultNominalThrustmN~mN; a scale factor of 0.9 gives \ResultEffectiveThrustmN~mN effective thrust. The stored GMAT configuration specifies $I_{\mathrm{sp}}=\ResultIsp$~s and disables propellant-mass decrement. Accordingly, propellant figures in this paper are analytical equivalents, rather than propagated fuel consumption. Table~\ref{tab:disposal} gives the deployment state, and Fig.~\ref{fig:design} shows the spacecraft CAD model.

\begin{table}[!htbp]
\centering
\caption{Spacecraft state at separation from the launch vehicle~\cite{cubequest2014,mathur2016low}.}
\label{tab:disposal}
\rowcolors{1}{}{gray!10}
\resizebox{\linewidth}{!}{%
\begin{tabular}{@{}cccccc@{}}
\toprule
Parameter & Value    & Unit   & Parameter & Value    & Unit  \\ \midrule
X         & $-$15015.4 & km   & SMA       & 205954.8 & km    \\
Y         & $-$23569   & km   & Ecc       & 0.9667   &       \\
Z         & 2241.505   & km   & Inc       & 28.6065  & deg   \\
VX        & $-$0.48554 & km/s & RAAN      & 65.9569  & deg   \\
VY        & $-$5.04876 & km/s & AoP       & 47.9162  & deg   \\
VZ        & $-$0.87999 & km/s & TA        & 122.4711 & deg   \\
Epoch     & \multicolumn{5}{c}{15 Dec 2017, 14:57:17.2 UTC}              \\ \bottomrule
\end{tabular}%
}
\end{table}

\subsection{Uncontrolled trajectory}

\begin{figure}[!htbp]
  \begin{subfigure}[b]{0.49\columnwidth}
    \includegraphics[
        width=\linewidth,
        height=\linewidth,
        keepaspectratio=false
    ]{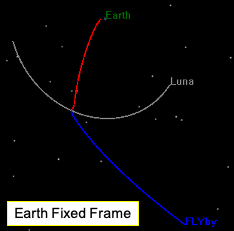}
    \caption{Earth-centered frame}
    \label{fig:1}
  \end{subfigure}
  \hfill
  \begin{subfigure}[b]{0.49\columnwidth}
    \includegraphics[
        width=\linewidth,
        height=\linewidth,
        keepaspectratio=false
    ]{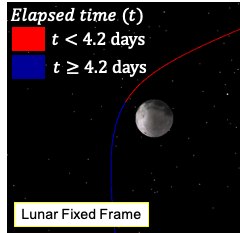}
    \caption{Moon-centered frame}
    \label{fig:2}
  \end{subfigure}
  \caption{Trajectory without actuation.}
  \label{fig:no_actuation}
\end{figure}

\begin{figure}[!htbp]
    \centering
    \includegraphics[width=0.49\textwidth]{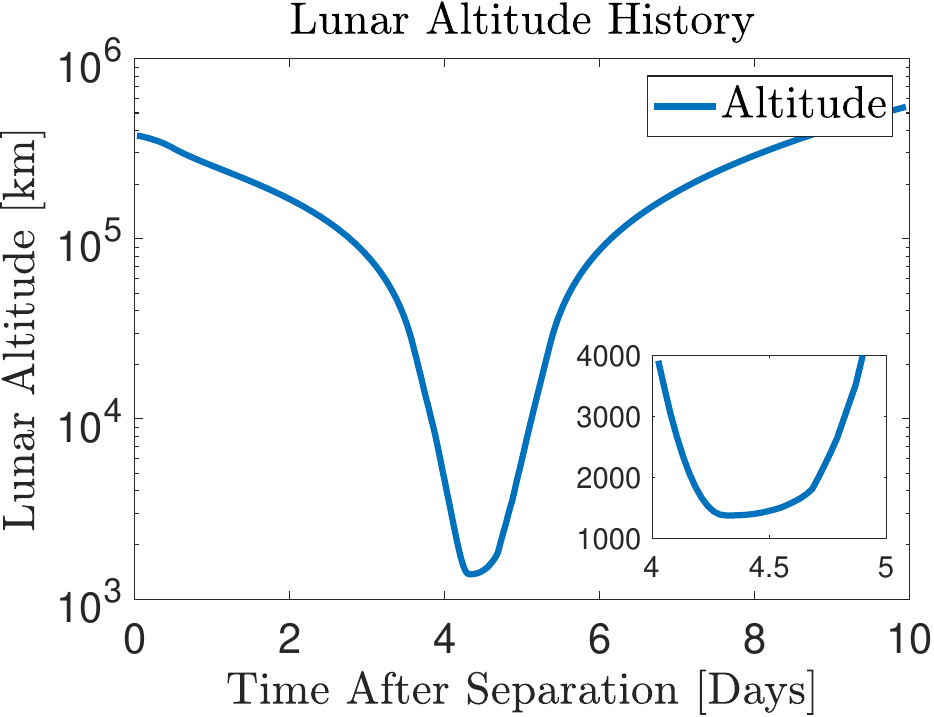}
    \caption{Lunar altitude history without actuation.}
    \label{fig:no_actuation_m}
\end{figure}

Ballistic propagation of the initial state reveals that the spacecraft performs an uncontrolled lunar flyby at approximately 1300~km altitude and subsequently escapes into a heliocentric orbit after $\sim$4.2~days (Figs.~\ref{fig:no_actuation}--\ref{fig:no_actuation_m}). To establish a stable lunar orbit, the mission trajectory is divided into four segments:

\textit{Segment~0: Detumbling and commissioning ($\sim$2~hours):} Immediately after separation, the spacecraft detumbles and establishes the communication network. This preliminary segment must complete before propulsive maneuvers can begin.

\textit{Phase~1: Pre-flyby deceleration:} Retrograde thrust is applied for \ResultPreFlybyDuration~days after the initial two-hour coast, modifying the subsequent Earth and Moon encounter sequence.

\textit{Phase~2: Lunar capture:} A transition burn, coast, and lunar capture burn produce a bound lunar orbit. At the end of the long thrust block containing capture, near day~\ResultCaptureEnd, the recorded osculating semi-major axis and eccentricity are approximately \ResultCaptureBlockSMA~km and \ResultCaptureBlockEcc, respectively.

\textit{Phase~3: Orbit circularization:} Repeated lunar-frame retrograde burns reduce the orbit size and eccentricity. The archived record contains \ResultCircularizationArcs~later lunar thrust blocks and terminates during the last recorded arc at day~\ResultMissionDays. The intended 100~km science orbit is a subsequent design target.

\subsection{Reference frames}

The reference GMAT propagators use the Moon as central body. States and trajectory projections are reported in two inertial frames:

\textit{Earth-Centered Inertial (ECI):} The origin is the Earth's center of mass, with $X$ toward the J2000 vernal equinox and $Z$ toward the north celestial pole. This frame is used for Earth-centered trajectory reports and Eq.~\eqref{eq:eom}.

\textit{Moon-Centered Inertial (MCI):} The origin is the Moon's center, with axes parallel to ECI J2000. Lunar orbital elements and the capture/circularization projections use this frame. Altitudes quoted here use a spherical lunar reference radius of 1738~km; endpoint apsis altitudes are $a(1\mp e)-R_{\mathrm{M}}$.

\subsection{Dynamical model}

The equations of motion are formulated in the ECI frame expressing the configured gravitational contributions in common coordinates:
\begin{equation}
\begin{aligned}
\ddot{\mathbf{r}}={}&-\frac{\mu_\oplus}{r^3}\mathbf{r}\\
&-\mu_\mathrm{M}\!\left(\frac{\mathbf{r}-\mathbf{r}_\mathrm{M}}{\|\mathbf{r}-\mathbf{r}_\mathrm{M}\|^3}+\frac{\mathbf{r}_\mathrm{M}}{r_\mathrm{M}^3}\right)\\
&+\mathbf{a}_\mathrm{3rd}+10^{-3}\frac{\mathbf{F}_\mathrm{thrust}}{m}.
\end{aligned}
\label{eq:eom}
\end{equation}
where $\mu_\oplus$ and $\mu_\mathrm{M}$ are the gravitational parameters of Earth and Moon, $\mathbf{r}_\mathrm{M}(t)$ is the lunar position from JPL ephemerides, and $\mathbf{a}_\mathrm{3rd}$ accounts for solar perturbations. The reference configuration uses Earth, Moon, and Sun point-mass gravity for the transfer. During lunar circularization it uses the LP165P lunar gravity field truncated to degree and order four, with Earth and Sun as third bodies. Solar radiation pressure, atmospheric drag, and relativistic corrections are disabled. 

Positions and velocities are expressed in km and km/s, whereas thrust is in N and mass in kg; the factor $10^{-3}$ converts thrust acceleration to km/s$^2$. The combined nominal thrust is bounded by:
\begin{equation}
\vectornorm{\mathbf{F}_\mathrm{total}} \leq \ResultNominalThrustmN~\text{mN}
\label{eq:thrust_total}
\end{equation}
with a 10\% inefficiency margin applied in the GMAT simulation (thrust scale factor 0.9), yielding an effective thrust of \ResultEffectiveThrustmN~mN.
The exhaust velocity is $v_e = g_0 I_\mathrm{sp} \approx \ResultExhaustKms$~km/s for the configured specific impulse.

\section{Trajectory Optimization}

\subsection{Problem formulation}

The capture problem is treated as finite-burn constrained targeting using the GMAT YUKON sequential quadratic programming solver~\cite{GMAT}. The fixed pre-flyby burn is outside the targeting block. The eight scalar targeting parameters comprise six thrust-direction components, the transition-burn duration, and the subsequent coast duration. Thrust directions are specified in Earth-centered and Moon-centered VNB frames for the transition and capture burns, respectively. The thruster coordinate frame defines the applied direction; it is distinct from the propagator's central-body selection. With these parameters collected in $\boldsymbol{\theta}$, the capture feasibility problem is
\begin{equation}
\begin{aligned}
\text{find}\quad & \boldsymbol{\theta} \\
\text{subject to}\quad & \dot{\mathbf{r}} = \mathbf{v}, \\
& \dot{\mathbf{v}} = \mathbf{a}_{\mathrm{grav}} + 10^{-3}\mathbf{F}_{\mathrm{thrust}}/m, \\
& 0 \leq \|\mathbf{F}_{\mathrm{thrust}}\| \leq F_{\max}, \\
& r_{\mathrm{perilune}}(t_f) \geq 6{,}000~\text{km}, \\
& C_3(t_f) \leq -0.11~\text{km}^2/\text{s}^2.
\end{aligned}
\label{eq:finite_burn_opt}
\end{equation}
The terminal point $t_f$ is a lunar periapsis event, so the position-vector magnitude (RMAG) at this event equals the perilune radius. The radius bound is measured from the Moon's center. In propagation, the effective thrust bound is $F_{\max}=\ResultEffectiveThrustmN$~mN after scaling. With $R_{\mathrm{M}}=1738$~km, a 6000~km perilune radius corresponds to 4262~km altitude. Here $C_3=v_{\mathrm{rel}}^2-2\mu_{\mathrm{M}}/\rho=-\mu_{\mathrm{M}}/a$, where $\rho=\|\mathbf{r}-\mathbf{r}_{\mathrm{M}}\|$ and $v_{\mathrm{rel}}$ is the lunar-relative speed. A negative value identifies a bound two-body osculating state; continued capture and surface clearance must also be checked along the propagated trajectory. The inequalities are terminal capture conditions and are not imposed as a 6000~km radius floor throughout circularization.

Table~\ref{tab:constraint_trade} distinguishes the capture conditions from the later circularization target. Eccentricity alone does not constrain orbit size or surface clearance. The capture block therefore combines energy and periapsis-radius conditions. The retained archive supports evaluation of the nominal trajectory, rather than a quantitative comparison of solver convergence rates or a completed robustness campaign.

\begin{table*}[!t]
\centering
\caption{Terminal capture conditions and the subsequent circularization target.}
\label{tab:constraint_trade}
\begin{tabular}{@{}lll@{}}
\toprule
Quantity & Condition & Application \\ \midrule
Lunar characteristic energy & $C_3\leq-0.11$~km$^2$/s$^2$ & Capture-burn endpoint \\
Perilune radius & $r_{\mathrm{perilune}}\geq6000$~km & Capture-burn endpoint \\
Eccentricity & $e<0.05$ & Later target; not reached in the record \\
Science-orbit altitude & 100~km & Later target; not reached in the record \\ \bottomrule
\end{tabular}
\end{table*}

\subsubsection{Lunar capture (Phase~2)}
After the pre-flyby burn, the recorded coast to the transition burn lasts \ResultPreTransitionCoast~days. The transition burn spans days \ResultTransitionStart--\ResultTransitionEnd~and lasts \ResultTransitionDuration~days. A \ResultTransitionCoast-day coast follows, then the lunar-frame thrust block containing capture spans days \ResultCaptureStart--\ResultCaptureEnd~and lasts approximately \ResultCaptureDuration~days. The first reported negative lunar $C_3$ occurs at day~\ResultCaptureDays; the preceding nonnegative sample brackets the change within days \ResultCaptureBracketDays. All later reported $C_3$ values remain negative to the end of the archived record. This sampling-based capture epoch is used consistently for the red dashed markers in the time histories.

\subsubsection{Orbit circularization (Phase~3)}
After capture, the Moon-centered thruster applies retrograde thrust in the lunar VNB frame. The configured repeating sequence thrusts from $\nu=240^\circ$ through periapsis to $\nu=130^\circ$, then coasts through apoapsis to $\nu=240^\circ$. Finite thrust changes both apsides over successive revolutions. The stored configuration requests up to 400 cycles, but the supplied thrust-state record resolves \ResultCircularizationArcs~later lunar thrust blocks, including the unfinished final block. Adjacent burns without a reported coast cannot be distinguished as separate events from this flag alone. The results below are restricted to this recorded interval.

Because SQP is a local method, the targeting sequence depends on its initialization. Impulsive reference solutions and manual timing refinement provide starting points for the finite-burn targeting. The archived run does not establish global optimality or quantify an iteration-count improvement over alternative initializations. A global search and a systematic convergence comparison remain future work.
\begin{figure}[!htbp]
    \centering
    \includegraphics[width=0.49\textwidth]{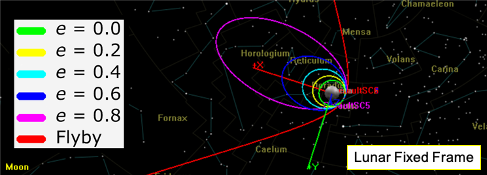}
    \caption{Impulsive-burn trajectories in the Moon-centered frame.}
    \label{fig:impulsive_lunar}
\end{figure}

\subsection{Impulsive-burn approximation}

Impulsive-burn reference solutions provide initial guesses for finite-burn targeting. Their endpoint assumptions differ from the recorded finite-burn solution, so their velocity increments are not treated as strict lower bounds for that solution. Table~\ref{table_impulsive} and Fig.~\ref{fig:impulsive_lunar} show the $\Delta$V requirements for different target eccentricities at lunar orbit insertion.

\begin{table}[!htbp]
\centering
\caption{Impulsive-burn $\Delta$V requirements.}
\rowcolors{1}{}{gray!10}
\begin{tabular}{@{}ll@{}}
\toprule
Eccentricity & $\Delta$V (m/s) \\ \midrule
0            & 977.8           \\
0.2          & 793.7           \\
0.4          & 639.3           \\
0.6          & 497.5           \\
0.8          & 364.5           \\ \bottomrule
\end{tabular}
\label{table_impulsive}
\end{table}

\subsection{Finite-burn optimization results}

The finite-burn results are evaluated from the archived GMAT position, velocity, distance, orbital-element, and thrust-state reports. Their elapsed-time columns are strictly increasing. Orbital elements extend to day~\ResultOrbitDaysPrecise~and the trajectory report to day~\ResultMissionDaysPrecise; both are reported as \ResultMissionDays~days at the precision used in this paper. No result beyond this interval is inferred.

At the recorded endpoint, the lunar osculating semi-major axis is \ResultSMAKm~km, eccentricity is \ResultEcc, inclination is \ResultInclinationDeg$^\circ$, and $C_3=\ResultCThree$~km$^2$/s$^2$. Using the same 1738~km reference radius as the lunar plots, the osculating perilune and apolune altitudes are approximately \ResultPeriluneKm~km and \ResultApoluneKm~km. Thus, the data demonstrate capture and partial circularization, but do not demonstrate the intended 100~km science orbit or the $e<0.05$ target.

\begin{figure*}[!p]
  \centering
  \captionsetup[subfigure]{font=footnotesize,skip=1pt}
  \begin{subfigure}[t]{0.48\textwidth}
    \centering\includegraphics[width=\linewidth]{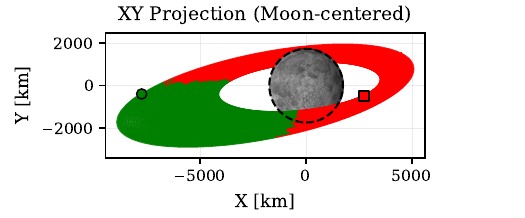}
    \caption{Moon-centered XY projection, last 10,000 points. Red indicates thrust active; green indicates coast.}
    \label{fig:finite_lunar_10k}
  \end{subfigure}\hfill
  \begin{subfigure}[t]{0.48\textwidth}
    \centering\includegraphics[width=\linewidth]{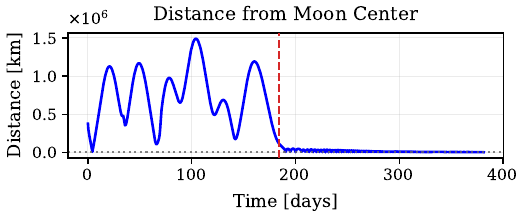}
    \caption{Distance from the Moon center.}\label{fig:moon_dist_full}
  \end{subfigure}
  \caption{Moon-centered trajectory and lunar-distance histories. The red dashed marker denotes capture at day~\ResultCaptureDays. Lunar texture: NASA/GSFC, shown as an illustrative disk.}
  \label{fig:lunar_panels}

  \vspace{4pt}

  \centering
  \captionsetup[subfigure]{font=footnotesize,skip=1pt}
  \begin{subfigure}[t]{0.48\textwidth}
    \centering\includegraphics[width=\linewidth]{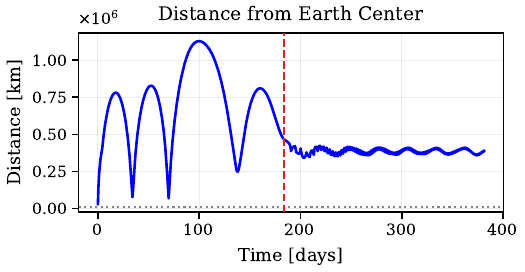}
    \caption{Distance from the Earth center.}\label{fig:earth_dist_full}
  \end{subfigure}\hfill
  \begin{subfigure}[t]{0.48\textwidth}
    \centering\includegraphics[width=\linewidth]{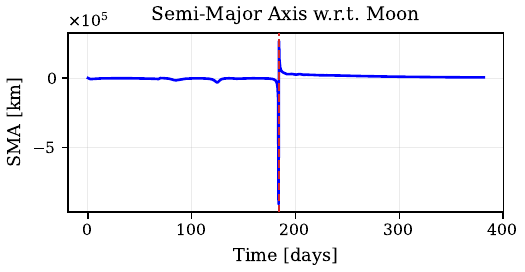}
    \caption{Lunar semi-major axis.}\label{fig:orb_sma_full}
  \end{subfigure}\par\vspace{2pt}
  \begin{subfigure}[t]{0.48\textwidth}
    \centering\includegraphics[width=\linewidth]{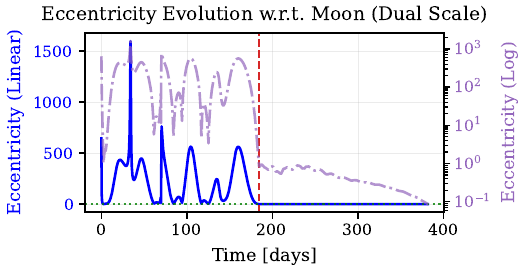}
    \caption{Lunar eccentricity, linear and logarithmic scales.}\label{fig:orb_ecc_full}
  \end{subfigure}\hfill
  \begin{subfigure}[t]{0.48\textwidth}
    \centering\includegraphics[width=\linewidth]{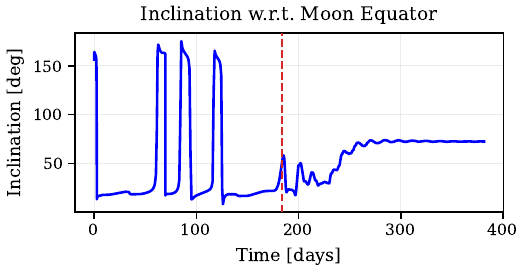}
    \caption{Inclination relative to the lunar equator.}\label{fig:orb_inc_full}
  \end{subfigure}\par\vspace{2pt}
  \begin{subfigure}[t]{0.48\textwidth}
    \centering\includegraphics[width=\linewidth]{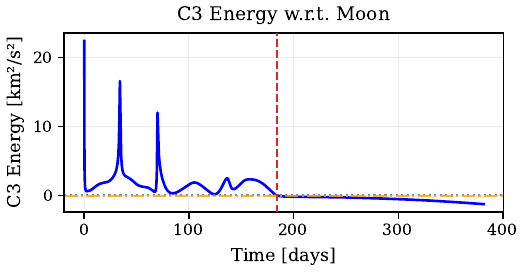}
    \caption{Lunar $C_3$ energy.}\label{fig:orb_c3_full}
  \end{subfigure}\hfill
  \begin{subfigure}[t]{0.48\textwidth}
    \centering\includegraphics[width=\linewidth]{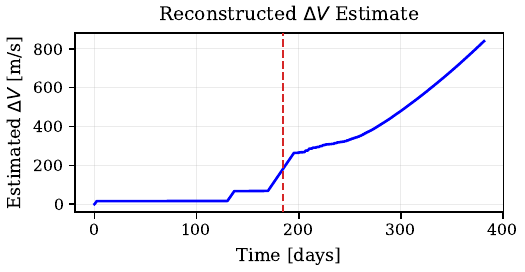}
    \caption{Reconstructed cumulative $\Delta V$ estimate.}\label{fig:deltav_full}
  \end{subfigure}
  \par\vspace{1pt}
  \includegraphics[width=0.98\textwidth]{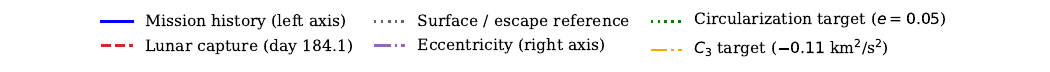}
  \caption{Mission history and lunar-orbit evolution. Red dashed lines mark the first recorded bound lunar state at day~\ResultCaptureDays~in all panels. Panels (a) through (f) show Earth distance, semi-major axis, eccentricity, inclination, lunar $C_3$, and reconstructed cumulative $\Delta V$. The shared history legend also applies to Fig.~\ref{fig:lunar_panels}(b).}
  \label{fig:mission_panels}
\end{figure*}

The cumulative-$\Delta V$ report field is zero throughout the archive. Therefore, Fig.~\ref{fig:mission_panels}(f) and Table~\ref{table_finite} use an explicitly reconstructed estimate,
\begin{equation}
\Delta V_{\mathrm{est}}=\frac{F_{\mathrm{eff}}}{m}\int_0^{t_{\mathrm{end}}}s(t)\,\mathrm{d}t,
\label{eq:dv_reconstruction}
\end{equation}
where $s(t)$ is the recorded binary thrust state and elapsed days are converted to seconds in the integral. Trapezoidal integration of the samples gives \ResultThrustDays~days of active thrust and an estimated \ResultDeltaV~m/s for \ResultEffectiveThrustmN~mN effective thrust at constant \ResultMassKg~kg mass. Switching times in Table~\ref{table_finite} are estimated at the midpoints between adjacent state samples. The table groups contiguous thrust-on samples; it does not identify internal GMAT burn-command boundaries. The estimates consequently inherit the report's sampling resolution.

For $I_{\mathrm{sp}}=\ResultIsp$~s, the corresponding ideal rocket-equation propellant equivalent is
\begin{equation}
 m_{p,\mathrm{eq}}=m_0\!\left[1-\exp\!\left(-\frac{\Delta V_{\mathrm{est}}}{g_0 I_{\mathrm{sp}}}\right)\right]
 \approx \ResultFuelKg~\mathrm{kg}.
\label{eq:fuel_equivalent}
\end{equation}
This is an analytical equivalent within the \ResultPropellantKg~kg allocation. It is not a simulated tank-depletion result or a validation of the hardware's achievable specific impulse; the propagation holds mass constant. A coupled mass-depleting simulation is required to establish an operational fuel margin.
\begin{figure}[!htbp]
  \centering
  \includegraphics[width=0.99\columnwidth]{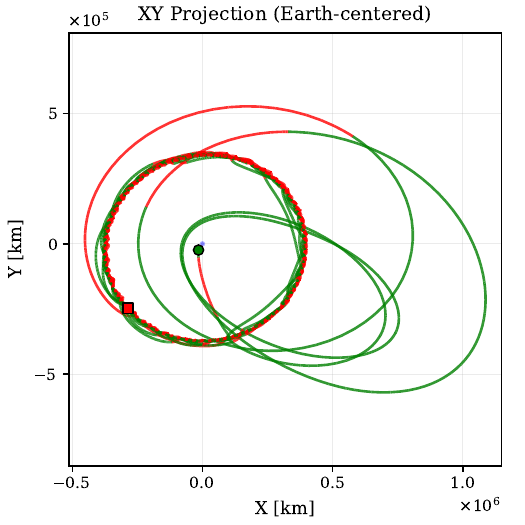}
  \caption{Earth-centered XY projection of the finite-burn trajectory (full resolution). Red: thrust active, green: coast.}
  \label{fig:finite_earth_full}
\end{figure}

\begin{figure*}[!t]
  \centering
  \begin{subfigure}[t]{0.48\textwidth}
    \centering
    \includegraphics[width=\linewidth]{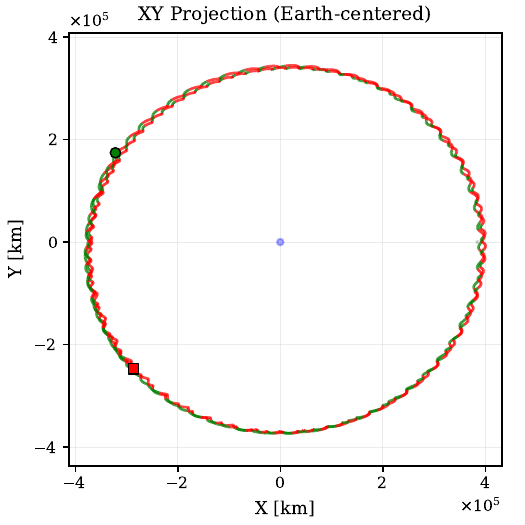}
    \caption{Earth-centered, last 10,000 points.}
    \label{fig:finite_earth_10k}
  \end{subfigure}\hfill
  \begin{subfigure}[t]{0.48\textwidth}
    \centering
    \includegraphics[width=\linewidth]{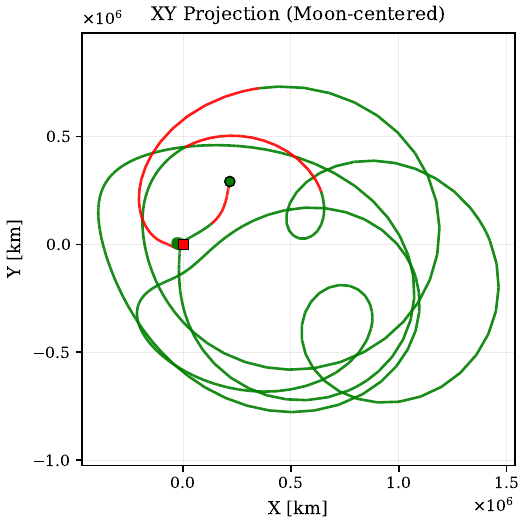}
    \caption{Moon-centered, full resolution.}
    \label{fig:finite_lunar_full}
  \end{subfigure}
  \par\vspace{1pt}
  \includegraphics[width=0.98\textwidth]{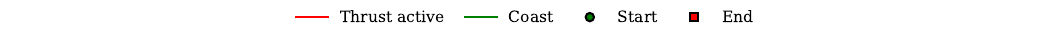}
  \caption{Earth-centered and Moon-centered projections of the finite-burn trajectory. The shared trajectory legend also applies to Figs.~\ref{fig:finite_earth_full} and \ref{fig:lunar_panels}(a).}
  \label{fig:trajectory_pair}
\end{figure*}

Figs.~\ref{fig:finite_earth_full}, \ref{fig:trajectory_pair}(a,b), and \ref{fig:finite_lunar_10k} show the Earth-centered and Moon-centered trajectory projections. Fig.~\ref{fig:mission_panels}(b--e) shows lunar semi-major axis, eccentricity, inclination, and characteristic energy. The positive-to-negative transition in $C_3$ identifies the first recorded bound lunar state at day~\ResultCaptureDays. Subsequent reductions in orbit size and eccentricity are visible through day~\ResultMissionDays. Earth and Moon distance histories appear in Figs.~\ref{fig:mission_panels}(a) and \ref{fig:moon_dist_full}, and the reconstructed velocity increment appears in Fig.~\ref{fig:mission_panels}(f). The green eccentricity reference and orange energy reference indicate the circularization target and capture-energy condition, respectively; the former remains unmet at the endpoint.

\begin{table}[!htbp]
\centering
\caption{Recorded thrust blocks and reconstructed velocity increments. The final block is unfinished.}
\label{table_finite}
\resizebox{\linewidth}{!}{\begin{tabular}{@{}lccc@{}}
\toprule
Maneuver & Epoch (days) & Thrust (days) & $\Delta V_{\mathrm{est}}$ (m/s) \\ \midrule
Pre-flyby burn & \ResultPreFlybyStart--\ResultPreFlybyEnd & \ResultPreFlybyDuration & \ResultPreFlybyDeltaV \\
Transition burn & \ResultTransitionStart--\ResultTransitionEnd & \ResultTransitionDuration & \ResultTransitionDeltaV \\
Capture/lunar thrust & \ResultCaptureStart--\ResultCaptureEnd & \ResultCaptureDuration & \ResultCaptureDeltaV \\
Later lunar thrust & \ResultCircularStart--\ResultMissionDays & \ResultCircularThrustDays & \ResultCircularDeltaV \\ \midrule
Total active thrust & & \ResultThrustDays & \ResultDeltaV \\
Total recorded mission & 0--\ResultMissionDays & & \\ \bottomrule
\end{tabular}}
\end{table}

\section{Limitations and Further Validation}

The archived result is a nominal trajectory calculation. It does not include a Monte Carlo campaign, navigation errors, thrust-direction errors, a hardware-failure assessment, or a quantified sensitivity ranking. No statistical success probability or operational robustness claim follows from this single record. In particular, negative osculating lunar energy throughout the reported post-capture interval does not establish indefinite stability after thrust termination.

The constant-mass approximation, prescribed thrust level, and configured specific impulse must be replaced by a validated propulsion and mass-depletion model before using the analytical propellant equivalent for mission budgeting. Solar radiation pressure and higher-degree lunar gravity should be included in a subsequent fidelity study. The remaining circularization to the intended 100~km science orbit and its operational lifetime must also be demonstrated explicitly.

\section{Conclusions}

The first reported bound lunar state occurs at day~\ResultCaptureDays. The record ends at day~\ResultMissionDays~with semi-major axis \ResultSMAKm~km and eccentricity \ResultEcc, corresponding to approximately \ResultPeriluneKm--\ResultApoluneKm~km altitude with the adopted lunar reference radius. The intended 100~km science orbit is not reached.

At constant \ResultMassKg~kg mass and \ResultEffectiveThrustmN~mN effective thrust, the reconstructed velocity increment is \ResultDeltaV~m/s over \ResultThrustDays~days of thrust. The ideal propellant equivalent is \ResultFuelKg~kg at the configured \ResultIsp~s specific impulse; depletion was not propagated. Completing the low-altitude transfer, validating the propulsion model, and assessing uncertainty remain future work.

\FloatBarrier


\begin{thebibliography}{10}

\bibitem{poghosyan2017cubesat}
A~Poghosyan and A~Golkar.
\newblock {CubeSat} evolution: Analyzing {CubeSat} capabilities for conducting
  science missions.
\newblock {\em Progress in Aerospace Sciences}, 88:59--83, 2017.

\bibitem{folta2016lunar}
D.~C. Folta, N.~Bosanac, A.~Cox, and K.~C. Howell.
\newblock The lunar {IceCube} mission design: Construction of feasible transfer
  trajectories with a constrained departure.
\newblock In {\em AAS/AIAA Spaceflight Mechanics Meeting}, 2016.
\newblock Paper AAS 16-285.

\bibitem{klesh2018marco}
A.~Klesh, B.~Clement, C.~Colley, J.~Essmiller, D.~Forgette, J.~Krajewski,
  A.~Marinan, and T.~Martin-Mur.
\newblock {MarCO}: Early operations of the first {CubeSats} to {Mars}.
\newblock In {\em 32nd Annual AIAA/USU Conference on Small Satellites}, 2018.

\bibitem{hardgrove2015lunar}
C.~Hardgrove, J.~Bell, J.~Thangavelautham, A.~Klesh, R.~Starr, T.~Colaprete,
  M.~Robinson, D.~Drake, et~al.
\newblock The lunar polar hydrogen mapper ({LunaH-Map}) mission: Mapping
  hydrogen distributions in permanently shadowed regions of the {Moon}'s south
  pole.
\newblock In {\em Annual Meeting of the Lunar Exploration Analysis Group},
  2015.
\newblock Abstract 2035.

\bibitem{cohen2015lunar}
B.~A. Cohen, P.~O. Hayne, B.~T. Greenhagen, and D.~A. Paige.
\newblock Lunar flashlight: Exploration and science at the {Moon} with a {6U
  CubeSat}.
\newblock In {\em AGU Fall Meeting Abstracts}, 2015.
\newblock Abstract EP52B-07.

\bibitem{orger2020horyu}
N.~C. {\"O}rger, M.~Cho, O.~B. Iskender, W.~S. Lim, A.~Chandran, K.~V. Ling,
  K.~H. Holden, C.~L. Chow, J.~Bellardo, P.~Faure, et~al.
\newblock {HORYU-VI}: International {CubeSat} mission to investigate lunar
  horizon glow.
\newblock In {\em Proceedings of the 71st International Astronautical
  Congress}. International Astronautical Federation, 2020.
\newblock Paper IAC-20-B4,2,7,x55547.

\bibitem{mccoy1974evidence}
J.~E. McCoy and D.~R. Criswell.
\newblock Evidence for a high altitude distribution of lunar dust.
\newblock In {\em Lunar and planetary science conference proceedings},
  volume~5, pages 2991--3005, 1974.

\bibitem{mccoy1976photometric}
J.~E. McCoy.
\newblock Photometric studies of light scattering above the lunar terminator
  from {Apollo} solar corona photography.
\newblock In {\em Lunar and planetary science conference proceedings},
  volume~7, pages 1087--1112, 1976.

\bibitem{rennilson1974surveyor}
J.~J. Rennilson and D.~R. Criswell.
\newblock Surveyor observations of lunar horizon-glow.
\newblock {\em The Moon}, 10(2):121--142, 1974.

\bibitem{criswell1973horizon}
D.~R. Criswell.
\newblock Horizon-glow and the motion of lunar dust.
\newblock In R.~J.~L. Grard, editor, {\em Photon and particle interactions with
  surfaces in space}, pages 545--556. Springer, 1973.

\bibitem{harris1972apollo}
R.~S. Harris, Jr.
\newblock {Apollo} experience report: Thermal design of {Apollo} lunar surface
  experiments package.
\newblock Technical Report NASA TN D-6738, NASA Manned Spacecraft Center, 1972.

\bibitem{gaier2005effects}
J.~R. Gaier.
\newblock The effects of lunar dust on {EVA} systems during the {Apollo}
  missions.
\newblock Technical Report NASA/TM-2005-213610, NASA Glenn Research Center,
  Cleveland, Ohio, 2005.

\bibitem{james2009pulmonary}
J.~T. James, C.~W. Lam, C.~Quan, W.~T. Wallace, and L.~Taylor.
\newblock Pulmonary toxicity of lunar highland dust.
\newblock Technical Report 2009-01-2379, SAE International, 2009.

\bibitem{linnarsson2012toxicity}
D~Linnarsson, J~Carpenter, B~Fubini, P~Gerde, L.~L. Karlsson, D.~J. Loftus,
  G.~K. Prisk, U~Staufer, E.~M. Tranfield, and W~van Westrenen.
\newblock Toxicity of lunar dust.
\newblock {\em Planetary and Space Science}, 74(1):57--71, 2012.

\bibitem{gaier2007effects}
J.~R. Gaier.
\newblock The effects of lunar dust on {EVA} systems during the {Apollo}
  missions.
\newblock Technical Report NASA/TM-2005-213610/REV1, NASA Glenn Research
  Center, Cleveland, Ohio, 2007.

\bibitem{o2011review}
B.~J. O'Brien.
\newblock Review of measurements of dust movements on the {Moon} during
  {Apollo}.
\newblock {\em Planetary and Space Science}, 59(14):1708--1726, 2011.

\bibitem{o2018paradigm}
B.~J. O'Brien.
\newblock Paradigm shifts about dust on the {Moon}: From {Apollo} 11 to
  chang'e-4.
\newblock {\em Planetary and Space Science}, 156:47--56, 2018.

\bibitem{severny1975measurements}
A.~B. Severny, E.~I. Terez, and A.~M. Zvereva.
\newblock The measurements of sky brightness on {Lunokhod-2}.
\newblock {\em The Moon}, 14(1):123--128, 1975.

\bibitem{glenar2011reanalysis}
D.~A. Glenar, T.~J. Stubbs, J.~E. McCoy, and R.~R. Vondrak.
\newblock A reanalysis of the {Apollo} light scattering observations, and
  implications for lunar exospheric dust.
\newblock {\em Planetary and Space Science}, 59(14):1695--1707, 2011.

\bibitem{colwell2007lunar}
J.~E. Colwell, S~Batiste, M~Hor{\'a}nyi, S~Robertson, and S~Sture.
\newblock Lunar surface: Dust dynamics and regolith mechanics.
\newblock {\em Reviews of Geophysics}, 45(2):RG2006, 2007.

\bibitem{feldman2014upper}
P.~D. Feldman, D.~A. Glenar, T.~J. Stubbs, K.~D. Retherford, G.~R. Gladstone,
  P.~F. Miles, T.~K. Greathouse, D.~E. Kaufmann, J.~W. Parker, and S.~A. Stern.
\newblock Upper limits for a lunar dust exosphere from far-ultraviolet
  spectroscopy by {LRO/LAMP}.
\newblock {\em Icarus}, 233:106--113, 2014.

\bibitem{barker2019searching}
M.~K. Barker, E~Mazarico, T.~P. McClanahan, X~Sun, G.~A. Neumann, D.~E. Smith,
  M.~T. Zuber, and J.~W. Head.
\newblock Searching for lunar horizon glow with the lunar orbiter laser
  altimeter.
\newblock {\em Journal of Geophysical Research: Planets}, 124(11):2728--2744,
  2019.

\bibitem{jpl2013}
J.~S. Parker and R.~L. Anderson.
\newblock {\em Low-Energy Lunar Trajectory Design}.
\newblock John Wiley \& Sons, Hoboken, New Jersey, 2014.

\bibitem{racca2009smart}
G.~D. Racca.
\newblock {SMART-1} from conception to {Moon} impact.
\newblock {\em Journal of Propulsion and Power}, 25(5):993--1002, 2009.

\bibitem{cubequest2014}
{NASA}.
\newblock {Cube Quest Challenge}.
\newblock \url{https://www.nasa.gov/image-article/cube-quest-challenge-2/},
  2015.
\newblock September 24, 2015.

\bibitem{mathur2016low}
R.~Mathur.
\newblock Low thrust trajectory design and optimization: Case study of a lunar
  {CubeSat} mission.
\newblock In {\em Proceedings of the 6th International Conference on
  Astrodynamics Tools and Techniques}, 2016.

\bibitem{GMAT}
S.~P. Hughes.
\newblock {General Mission Analysis Tool (GMAT): Application to GSFC Mission
  Design}.
\newblock NASA Goddard Space Flight Center presentation, March 2016.

\end{thebibliography}
\end{document}